\documentclass[english,prb,pop]{revtex4-1}
\usepackage[T1]{fontenc}
\usepackage[latin9]{inputenc}
\usepackage{color}
\usepackage{bm}
\usepackage{graphicx}
\usepackage{esint}

\makeatletter
\usepackage{lineno}

\makeatother

\usepackage{babel}
\begin{document}

\title{Dust levitation in an inverse sheath}

\author{Rinku Deka and Madhurjya P Bora}

\affiliation{Physics Department, Gauhati University, Guwahati 781014, India}
\email{mpbora@gauhati.ac.in}

\begin{abstract}
The results of an analysis of the physics of levitation of negatively
charged dust particles over a surface (wall) in an inverse sheath
are reported. It is shown that under suitable parameter regime, the
ion-drag force may balance the combined electrostatic and gravitational
force on the dust particles owing to its hollow profile as one moves
away from the surface. Our analysis shows that the parameter regimes
in which such a situation may result is realizable in laboratory and
space plasma environments, particularly the near-surface dayside lunar
plasma. The lunar surface and dust grains are electrostatically charged
due to the interaction with the solar wind plasma environment and
the photoemission of electrons due to solar UV radiation. This results
in a process that charges the surface positively and generates a near-surface
photoelectron inverse plasma sheath. The potential structure changes
from a monotonic classical sheath to an inverse sheath as the emitted
electron density becomes larger than the plasma electron density.
In a relatively newer, recently developed charging model, called the
Patched Charge Model, it has shown both theoretically and experimentally
that even in photoelectron-rich environment, dust particles lying
on a regolith surface can attain large negative potential due to formation
of micro cavities. This negative potential may reach such values so
that dust mobilization and lofting may become possible. In our work,
we have assumed the existence of such negatively charged dust particles
in a photoelectron-rich environment and argue that once the dust lofting
is effected, the levitation can be sustained through the ion-drag
force. The conditions of levitation are investigated for these dust
particles and the levitation distances from the lunar surface are
calculated.
\end{abstract}
\maketitle

\section{Introduction}

The subject of plasma-wall interaction has always been an intriguing
one. The plasma sheath, which usually forms in the vicinity of a surface,
is a site of intense nonlinear activities. In general, the wall becomes
negatively charged and a classical plasma sheath is formed \citep{1}.
However, when there are emissions from wall such as secondary electron
emission and photoemission, the wall becomes positively charged and
an inverse sheath is formed \citep{2}. In this work, we discuss the
dynamics of negatively charged dust particles in such an inverse sheath.
As an example case, we take the case of lunar plasma sheath, though
the work can be relevant in other laboratory and space plasmas, as
well.

Devoid of any detectable atmosphere, the lunar surface is constantly
in interaction with the solar wind plasma and solar ultraviolet (UV)
radiation and it serves as an ideal natural environment for studying
plasma-wall interaction. On the sunlit side, the lunar surface collects
solar wind ions and electrons as well as is also subjected to photoemission.
Primarily due to the solar UV photons, photoelectrons are emitted
from the lunar surface and the surface potential becomes positive.
These photoelectrons create a sheath above the dayside lunar surface
with a barrier of negative potential \citep{3}. If the ratio of the
emitted electrons to the collected electrons is small, then the surface
potential is negative relative to the plasma. In this case, a monotonic
classical sheath is formed. When the emitted electron density becomes
larger than the collected electron density, the potential structure
becomes a monotonic inverse sheath, where the surface potential is
more positive than the ambient plasma potential \citep{2}. Some of
the early works towards understanding of the sheath dynamics include
the works of Hobbs and Wesson \citep{4}, who made a fundamental fluid
theory for sheath in the presence of emitted electrons in case of
a planar surface which was immersed in a plasma that consists of cold
ions and Maxwellian electrons. Works of many authors \citep{5} contributed
toward significant analytical insights to the photoelectron sheath
over the lunar surface. Further computational studies \citep{6,7}
include results from Particle-In-Cell (PIC) simulations of photoelectron
sheath on lunar surface under different conditions.

So far as the investigation about dust-plasma interactions in laboratory
plasmas are concerned, there are quite a number of important works,
which have contributed to our understanding about this \citep{Bailung}.
On the experimental front, both space borne and laboratory experiments
provided us with valuable information about the potentials in the
electron-emitting sheaths \citep{8,9,10}. Some recent works which
are devoted to dusty plasma physics in the near-surface lunar layer,
which demonstrate that in the terminator region, a sheath-like plasma
layer exists, resulting electric field which can cause lofting of
dust particles of size $\sim2-3\,{\rm \mu m}$ to about $30\,{\rm cm}$
\citep{Popelb}. A very recent work investigates the \emph{absence
of dead zone} (where dust particles can not rise) near the lunar latitude
$\sim80^{\circ}$ \citep{Popela}.

Coming back to the issue of dust dynamics in a plasma sheath, we note
that dust particles are ubiquitous to space and laboratory plasmas
and the lunar plasma sheath is also no exception. Bombardment of the
lunar surface by high energetic electrons and ions causes ejection
of dust particles from the lunar regolith to the plasma sheath. The
movement of active dust transport near surfaces of airless bodies
in the solar system including dust grains levitated above the lunar
surface was first observed by the Surveyor spacecraft and the Apollo
missions \citep{11,12,13,14}. The dust particles are abundant in
all kinds of plasmas and it changes the plasma dynamics by being a
constituent plasma component. From various satellite observations
like Lunar Prospectror (LP) \citep{lunar_prospector} and Apollo-Era
Missions \citep{apollo1,apollo2,apollo3}, we came to know about the
existence of a complex as well as coupled plasma and layers of dust
particles above the lunar surface \citep{12,15}. These micron and
sub-micron sized dust grains from the lunar regolith get charged due
to collection of electrons and ions from the plasma, photoemission,
and secondary electron emission. In some recent works \citep{gcdas,16},
the charge balance over the lunar surface and levitation of dust particles
along with the steady state altitude dependence of dust density and
particle size in the case of classical sheath have been investigated.

In this work, we present an analysis of the dynamics of \emph{negatively
charged }dust particles in an inverse sheath. As mentioned above,
an inverse sheath usually forms on the dayside of the lunar surface
due to photoemission. It is customary to assume that the dust particles
in an inverse sheath also charge to positive potential due to the
dust photoemission. While this is mostly true, there are situations
where dust particles in an inverse sheath can also charge to high
negative potential. One of the newest model on dust charging which
has been developed is the \emph{Patched Charge Model }(PCM), which
can account for existence for dust particles with large negative charge
in an environment which supports emission and absorption of photoelectrons.
An example such a situation is the lunar plasma environment and lunar
horizon glow \citep{3} is such a phenomena which may have an explanation
through the PCM \citep{Wang,Schwan,19}. In our work, we consider
such a situation, where we have self-consistently calculated the dust-charge
from the current-balance equations. Our analysis shows that the ion-drag
force on the dust particles can sustain a levitation of even negatively
charged particles in an inverse sheath and the combined effect of
electrostatic, gravity, and ion-drag force can cause the dust particles
of various sizes to levitate \citep{17} and form\emph{ bands} above
the lunar surface, the heights of which compare very well with the
observational data \citep{14,18}. We believe that this is the first
time that the possibility of levitation of negatively charged dust
particles in an inverse sheath is reported.

In Sec II, we have considered our model and derived the electron and
photoelectron density assuming both of them as Maxwellian. In Sec
III, we have developed the potential structure and sheath equation.
In Sec IV, we have considered the dynamics of the dust particles in
the inverse sheath regime, where charging processes and current balanced
are self-consistently included. In this section, we have also calculated
the various forces acting on a negatively charged dust grain when
it is in the inverse sheath. We have also calculated the dust sizes
with respect to the distances from the surface. In Sec V, we summarize
our results.

\section{Plasma model}

We consider a one dimensional collisionless inverse plasma sheath
consisting of electrons and ions with considerable presence of dust
particles. As our region of interest is ion-acoustic time scale, the
dust particles do not take part in the plasma dynamics.\emph{ }In
1-D, the basic equations are ion continuity, ion momentum, and Poisson's
equation,
\begin{eqnarray}
\frac{\partial n_{i}}{\partial t}+\frac{\partial}{\partial x}(n_{i}u_{i}) & = & 0,\label{eq:cont-1}\\
\frac{\partial u_{i}}{\partial t}+u_{i}\frac{\partial u_{i}}{\partial x} & = & -\frac{1}{m_{i}n_{i}}\frac{\partial p_{i}}{\partial x}-\frac{e}{m_{i}}\frac{\partial\phi}{\partial x}.\label{eq:mom-1}
\end{eqnarray}
The two populations of electrons --- the plasma electrons (density
$n_{{\rm pe}}$) and the photoelectrons (density $n_{{\rm ph}}$)
emitted by the wall, are assumed to be Maxwellian, described by their
respective distributions \citep{5},
\begin{eqnarray}
f_{{\rm pe}}(x,\bm{v}) & = & n_{{\rm pe}0}\left(\frac{m_{e}}{2\pi T_{e}}\right)^{3/2}e^{-(m_{e}v^{2}/2-e\phi)/T_{e}}\\
f_{{\rm ph}}(x,\bm{v}) & = & n_{w}\left(\frac{m_{e}}{2\pi T_{{\rm ph}}}\right)^{3/2}e^{-m_{e}v^{2}/(2T_{{\rm ph}})+e(\phi-\phi_{w})/T_{{\rm ph}}},
\end{eqnarray}
where $T_{e,{\rm ph}}$ are electron and photoelectron temperatures
(expressed in energy units), $\phi$ is the plasma potential, and
$\phi_{w}$ is the potential at the wall. For an inverse sheath, the
potential $\phi_{w}>0$ and monotonically reduces to zero at the bulk
plasma. We assume that far away from the wall, the plasma quasi-neutrality
is maintained by the plasma electrons, photoelectrons, ions, and the
charged dust particles,
\begin{equation}
n_{e0}+z_{d0}n_{d}=n_{i0},\qquad n_{{\rm pe0}}+n_{{\rm ph0}}=n_{e0},
\end{equation}
where the subscript `$0$' denotes bulk plasma values. While the dust
density is \emph{not }a dynamic quantity but the dust charge number
$z_{d}$ is and $z_{d0}$ denotes its value in the bulk plasma away
from the wall. Both plasma electron and photoelectron densities can
be assumed to be close to Boltzmannian. While the plasma electron
is described by a single population, photoelectron density is described
by two populations --- a sheath-limited population and one which
contribute to the bulk plasma electrons,
\begin{eqnarray}
n_{{\rm pe}} & = & \int_{-\infty}^{\infty}\int_{-\infty}^{\infty}\int_{\sqrt{2e\phi/m_{e}}}^{\infty}f_{{\rm pe}}(x,\bm{v})\,d\bm{v}\nonumber \\
 & = & \frac{1}{2}n_{{\rm pe}0}\,e^{e\phi/T_{e}}\,{\rm erfc}\left(\frac{e\phi}{T_{e}}\right)^{1/2},\label{eq:npe-1}\\
n_{{\rm ph}} & = & \int_{-\infty}^{\infty}\int_{-\infty}^{\infty}\int_{\sqrt{2e\phi/m_{e}}}^{\infty}f_{{\rm ph}}(x,\bm{v})\,d\bm{v}+2\int_{-\infty}^{\infty}\int_{-\infty}^{\infty}\int_{0}^{\sqrt{2e\phi/m_{e}}}f_{{\rm ph}}(x,\bm{v})\,d\bm{v}\nonumber \\
 & = & \frac{1}{2}n_{w}\,e^{e(\phi-\phi_{w})/T_{{\rm ph}}}\left[1+{\rm erf}\left(\frac{e\phi}{T_{{\rm ph}}}\right)^{1/2}\right]=\frac{1}{2}n_{{\rm phw}}\,e^{e\phi/T_{{\rm ph}}}\left[1+{\rm erf}\left(\frac{e\phi}{T_{{\rm ph}}}\right)^{1/2}\right],\label{eq:nph-1}
\end{eqnarray}
where $n_{w}$ is the photoelectron density at the wall and is related
to the bulk photoelectron density through the relation $n_{{\rm phw}}=n_{w}e^{-e\phi_{w}/T_{{\rm ph}}}$.
Note that the photoelectron density moving away from the wall (when
they are produced) is half of the total photoelctrons produced which
contribute to the bulk plasma. Similarly, the plasma electron density
moving toward the wall (far away from the wall, at $\infty$) is half
of the bulk electrons at $\infty$. The ions are assumed to be polytropic,
\begin{equation}
p_{i}\propto n_{i}^{\gamma}.
\end{equation}
However, as the temperature remains constant, in what follows, we
assume $\gamma=1$. The model is closed by the Poisson equation,
\begin{equation}
\epsilon_{0}\frac{\partial^{2}\phi}{\partial x^{2}}=e(n_{{\rm pe}}+n_{{\rm ph}}-n_{i}+z_{d}n_{d}).
\end{equation}
Note that the dust density $n_{d}$ is not a dynamical variable. We
choose to normalize the ion density by the equilibrium values $n_{i0}\equiv n_{0}$,
the plasma potential by $T_{e}/e$, length by Debye length $\lambda_{D}$,
velocity by ion-thermal velocity $u_{s}=\sqrt{T_{e}/m_{i}}$, and
time by $\lambda_{D}/u_{s}$. So, Eqs.(\ref{eq:cont-1},\ref{eq:mom-1})
and (\ref{eq:npe-1},\ref{eq:nph-1}) can be normalized as, 
\begin{eqnarray}
\frac{\partial n_{i}}{\partial t}+\frac{\partial}{\partial x}(n_{i}u_{i}) & = & 0,\label{eq:cont}\\
\frac{\partial u_{i}}{\partial t}+u_{i}\frac{\partial u_{i}}{\partial x}+\frac{\sigma}{n_{i}}\frac{\partial p_{i}}{\partial x} & = & -\frac{\partial\phi}{\partial x},\label{eq:mom}\\
n_{{\rm pe}} & = & \frac{1}{2}\delta_{{\rm pe}}e^{\phi}\,{\rm erfc}\,(\phi^{1/2}),\label{eq:npe}\\
n_{{\rm ph}} & = & \frac{1}{2}\delta_{{\rm ph}}\,e^{\phi/\sigma_{{\rm ph}}}\left[1+{\rm erf}\left(\frac{\phi}{\sigma_{{\rm ph}}}\right)^{1/2}\right],\label{eq:nph}
\end{eqnarray}
where $n_{{\rm pe},{\rm ph}}$ are densities of the plasma electrons
and photoelectrons normalized by the equilibrium electron density
$n_{e0}$, $\delta_{{\rm pe,ph}}=n_{{\rm pe0,ph0}}/n_{e0}$ is a measure
of the fraction of plasma and photoelectron densities to that of total
bulk plasma electron density. The electron density at any instant
is $n_{e}=n_{{\rm pe}}+n_{{\rm ph}}$. Note that $(\delta_{{\rm pe}}+\delta_{{\rm ph}})/2=1$.
The normalized Poisson equation is given by,
\[
\frac{\partial^{2}\phi}{\partial x^{2}}=n_{{\rm pe}}+n_{{\rm ph}}-\delta_{i}n_{i}+\delta_{d}z_{d},
\]
where $\delta_{i,d}=(n_{0},z_{d0}n_{d})/n_{e0}$ are fractions of
ion and dust densities to the bulk plasma electron density. The dust
charge number is expressed in terms of its value in the bulk plasma.
The ion pressure is normalized to $n_{0}T_{i}$ and $\sigma=T_{i}/T_{e}$.

\section{The potential structure and sheath equation}

Far away from the boundary, the plasma potential vanishes i.e.\ as
$x\to\infty$, $\phi\to0,u\to M,p_{i}\to1,n_{i}\to1,z_{d}\rightarrow1$,
where $M$ is the mach number. In a co-moving frame of the wave, we
have from Eqs.(\ref{eq:cont},\ref{eq:mom}),
\begin{eqnarray}
\frac{\partial}{\partial x}(n_{i}u_{i}) & = & 0,\\
-u_{i}\frac{\partial u_{i}}{\partial x}+\frac{\sigma}{n_{i}}\frac{\partial p_{i}}{\partial x} & = & -\frac{\partial\phi}{\partial x},\label{eq:mom1}
\end{eqnarray}
From the continuity equation and integrating the Eq.(\ref{eq:mom1})
we have,
\begin{equation}
n_{i}=\frac{M}{\sqrt{\sigma W(z)}},\qquad z=\frac{M^{2}}{\sigma}\exp\left(\frac{M^{2}+2\phi}{\sigma}\right),\label{eq:phi}
\end{equation}
where $W(z)$ is the Lambert-W function and we have used the relation
$u_{i}=M/n_{i}$. The dependence of $n_{i}$ with the distance from
the wall is shown in Fig.\ref{fig:The-potential-profile}. Integrating
the Poisson's equation, we have 
\begin{equation}
\frac{1}{2}\left(\frac{d\phi}{d\xi}\right)^{2}+V(\phi,M,\sigma,\delta_{i},\sigma_{{\rm ph}},\delta_{{\rm ph}})=0,\label{eq:sheath equation}
\end{equation}
where $V$ is the Sagdeev or pseudo potential \citep{Sagdeev}, 
\begin{eqnarray}
V\left(\phi,M,\sigma,\delta_{i},\sigma_{{\rm ph}},\delta_{{\rm ph}}\right) & = & \left(1-2\sqrt{\phi/\pi}-e^{\phi}{\rm erfc}\sqrt{\phi}\right)\left(1-\frac{1}{2}\delta_{{\rm ph}}\right)+\frac{1}{2}\delta_{{\rm ph}}\sigma_{{\rm ph}}\,\xi(\phi)\nonumber \\
 &  & +\,\delta_{i}\int_{0}^{\phi}n_{i}(\phi)\,d\phi+(1-\delta_{i})\int_{0}^{\phi}z_{d}(\phi)\,d\phi,
\end{eqnarray}
and
\begin{equation}
\xi(\phi)=1-e^{\phi/\sigma_{{\rm ph}}}\left(1+{\rm erf}\sqrt{\phi/\sigma_{{\rm ph}}}\right)+2\left(\frac{\phi}{\pi\sigma_{{\rm ph}}}\right)^{1/2},
\end{equation}
The pseudo potential $V$ satisfies the boundary condition $V|_{\phi=0}=0$.
Besides, $V$ must be negative for all $\phi$ for a physically viable
potential profile.

\subsection{Wall potential}

\begin{figure}
\begin{centering}
\includegraphics[width=0.5\textwidth]{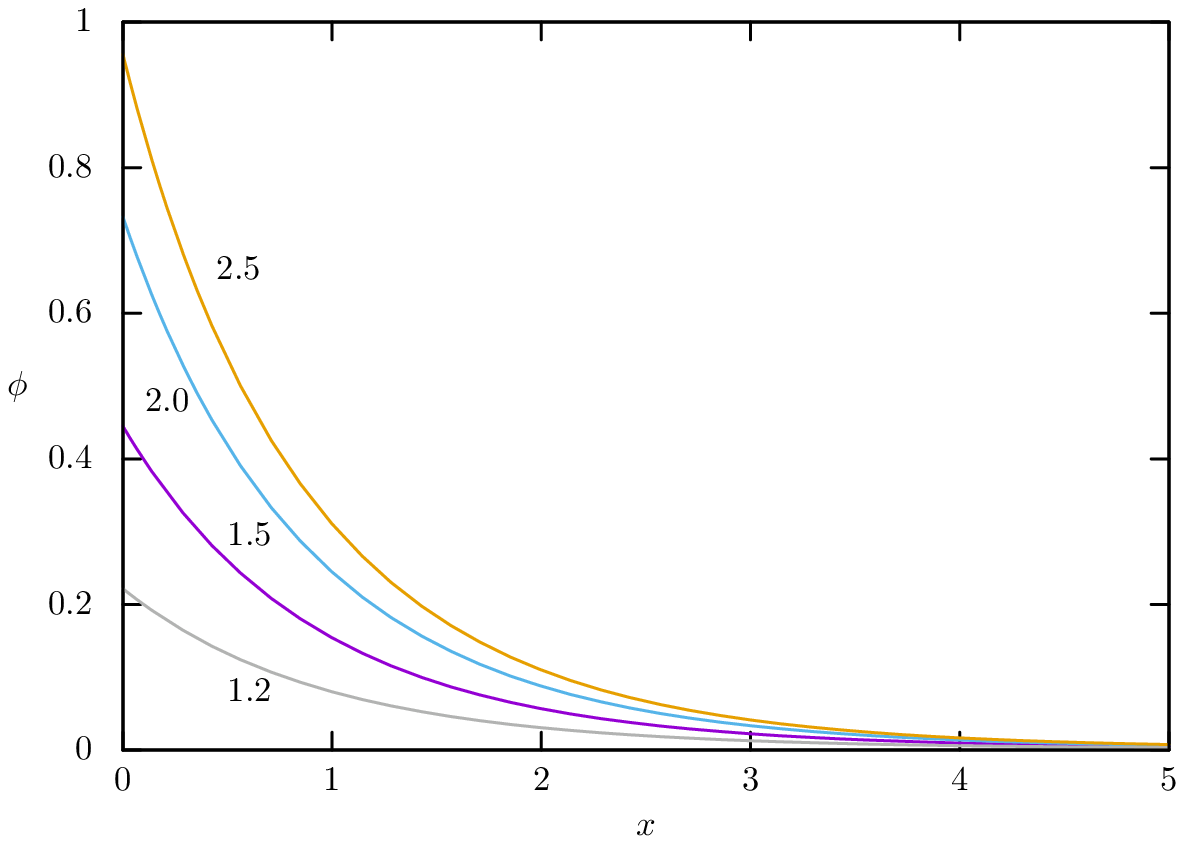}\hfill{}\includegraphics[width=0.5\textwidth]{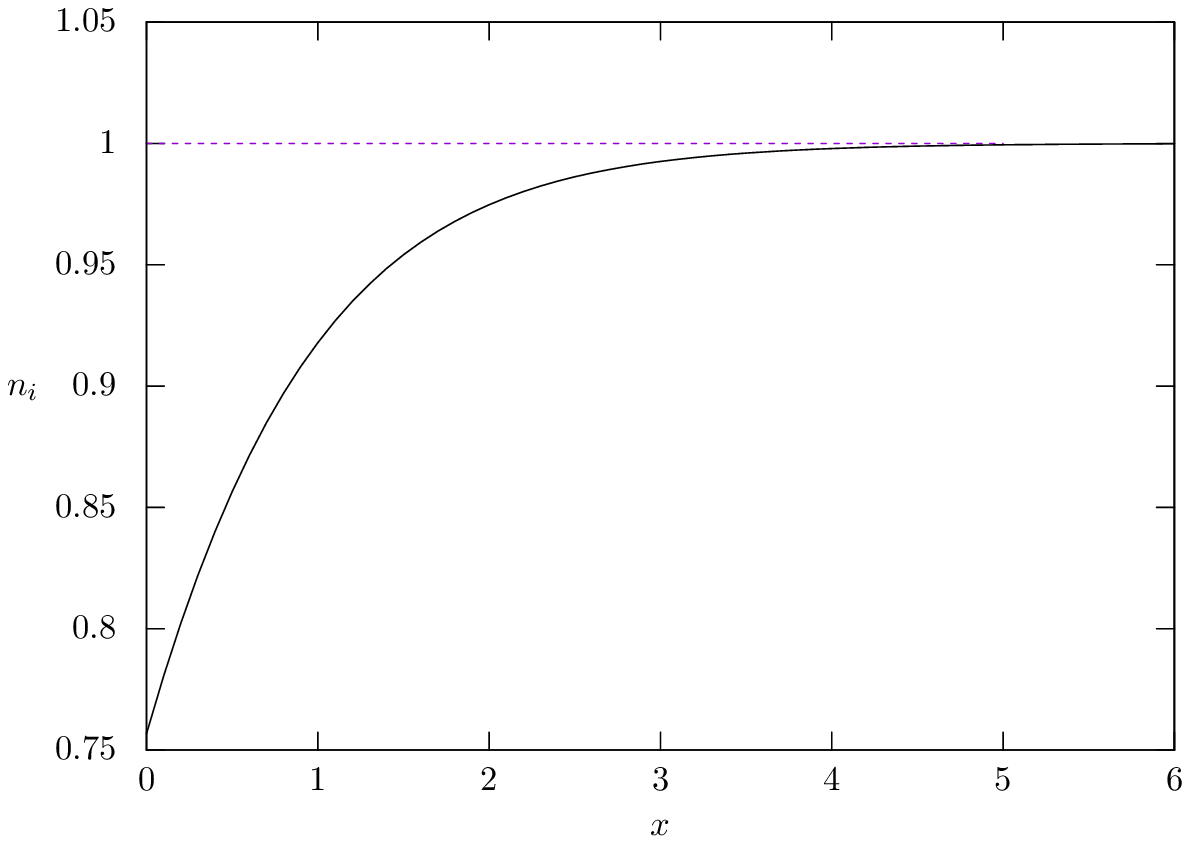}
\par\end{centering}
\caption{\label{fig:The-potential-profile}The potential profile (left) showing
the formation of inverse sheath (the wall is at $x=0$). The parameters
used are $n_{e0}=10^{12}\,{\rm m}^{-3},T_{e}=1\,{\rm keV},J_{h\nu}=4.5\,\mu{\rm A/m^{2}},M=1.2,\sigma=1,\delta_{i}=1.1,\sigma_{{\rm ph}}=1$.
The numbers in the figure indicates $\delta_{{\rm ph}}$. The panel
on the right shows the normalized ion density as one moves from the
wall. All the parameters are same with \textcolor{red}{$\delta_{{\rm ph}}=2.5$.}}
\end{figure}

To determine the wall potential, we employ the boundary condition
at infinity (at a distance far away from the wall) where the fluxes
for primary electrons, ions, and secondary electrons balance to have
the net plasma current density zero. In our case the secondary electrons
are the photoelectrons emitted by the wall,
\begin{equation}
J_{{\rm pe}}+J_{i}+J_{{\rm ph}}=0,\label{eq:wall-balance}
\end{equation}
where
\begin{eqnarray}
J_{{\rm pe}} & = & e\int_{-\infty}^{\infty}\int_{-\infty}^{\infty}\int_{-\infty}^{0}f_{{\rm pe}}(\infty,\bm{v})\,d\bm{v}=-en_{{\rm pe0}}\left(\frac{T_{e}}{2\pi m_{e}}\right)^{1/2},\\
J_{{\rm ph}} & = & e\int_{-\infty}^{\infty}\int_{-\infty}^{\infty}\int_{\sqrt{2e\phi_{w}/m_{e}}}^{\infty}f_{{\rm ph}}(0,\bm{v})\,d\bm{v}=en_{w}\left(\frac{T_{{\rm ph}}}{2\pi m_{e}}\right)^{1/2}e^{-e\phi_{w}/T_{{\rm ph}}},\\
J_{i} & = & en_{0}M\left(\frac{T_{e}}{m_{i}}\right)^{1/2}.
\end{eqnarray}
Note that the sheath-limited photoelectrons do not contribute to the
net plasma current. Eq.(\ref{eq:wall-balance}) can be written in
a normalized form as
\begin{equation}
\phi_{w}=\sigma_{{\rm ph}}\,\ln\left[\delta_{m}\delta_{w}\left(\frac{1+\sqrt{\sigma_{{\rm ph}}}}{2\delta_{m}-\delta_{i}M\sqrt{2\pi}}\right)\right]=\sigma_{{\rm ph}}\,\ln\left(\frac{\delta_{w}}{\delta_{{\rm ph}}}\right)
\end{equation}
which determines the wall potential $\phi_{w}$. $\delta_{m}=\sqrt{m_{i}/m_{e}}\approx43$.
In Fig.\ref{fig:The-potential-profile}, we show the structure of
the plasma potential $\phi$ which shows the formation of inverse
sheath. In the figure, the wall is at $x=0$.

\section{Dusts in plasma sheath}

\subsection{Current balance}

We now consider various currents to the surface of the dust grains
which are due to electrons $(I_{e})$, ions $(I_{i})$, photoelectrons
from the wall $(I_{{\rm ph}})$, and the photoemission current emitted
by the dust grain itself $(I_{h\nu})$. We assume that the currents
to the surface of the dust particles remain balanced all throughout,
\begin{equation}
\sum I=0=I_{e}+I_{i}+I_{{\rm ph}}+I_{h\nu}.
\end{equation}
This can be justified in the parameter space where this analysis is
relevant. For negatively charged dust particles, the expressions for
the currents are given by \citep{Shukla},
\begin{equation}
I_{e}=-4\pi r_{d}^{2}en_{{\rm pe}}\left(\frac{T_{e}}{2\pi m_{e}}\right)^{1/2}e^{e\phi_{d}/T_{e}},
\end{equation}
where $\phi_{d}$ is the dust potential and $r_{d}$ is the average
radius of a dust particle,
\begin{eqnarray}
I_{i} & = & 4\pi r_{d}^{2}en_{i}\left(\frac{T_{i}}{2\pi m_{i}}\right)^{1/2}\left(1-\frac{e\phi_{d}}{T_{i}}\right),\\
I_{h\nu} & = & \pi r_{d}^{2}J_{h\nu},
\end{eqnarray}
where $J_{h\nu}=eJ_{p}Q_{{\rm ab}}Y_{p}$ is the photoemission current
density. Here $J_{p}$ is the photon flux, $Q_{{\rm ab}}$ is the
efficiency of absorption for photons which is $\sim1$, and $Y_{p}$
is the photoelectron yield. For lunar dust $J_{h\nu}\sim4.5\,\mu{\rm A}/{\rm m}^{2}$
\citep{Feuerbacher},
\begin{equation}
I_{{\rm ph}}=-4\pi r_{d}^{2}en_{{\rm ph}}\left(\frac{T_{{\rm ph}}}{2\pi m_{e}}\right)^{1/2}e^{e\phi_{d}/T_{{\rm ph}}}.
\end{equation}
Using the expression for $n_{{\rm pe},{\rm ph}}$ from Eqs.(\ref{eq:npe},\ref{eq:nph}),
the normalized current balance equation can be written as,
\begin{equation}
-\delta_{{\rm pe}}\delta_{m}e^{\phi+\phi_{d}}\,{\rm erfc}\,(\phi^{1/2})+2n_{i}\delta_{i}\sqrt{\sigma}\left(1-\frac{\phi_{d}}{\sigma}\right)+\sqrt{\pi/2}\tilde{J}_{h\nu}-\delta_{{\rm ph}}\delta_{m}e^{(\phi+\phi_{d})/\sigma_{{\rm ph}}}\left[1+{\rm erf}\left(\frac{\phi}{\sigma_{{\rm ph}}}\right)^{1/2}\right]=0,\label{eq:dcharge}
\end{equation}
where $\tilde{J}_{h\nu}$ is the normalized photoemission current
density. Note that the dust charge $Q_{d}=q_{d}z_{d}$ is related
to the relative dust potential $\phi_{d}$ as (assuming a spherical
dust grain) $Q_{d}\equiv C\phi_{d}=4\pi\epsilon_{0}r_{d}\phi_{d},\phi_{d}=\phi_{g}-\phi,$
where $C$ is the capacitance of the spherical dust grain and $\phi_{g}$
is the absolute grain potential (with respect to zero). The dust charge
number $z_{d}$, thus can be expressed in terms of the dust potential
$\phi_{d}$ relative to the dust charge number at zero potential $z_{d}(\phi)=\phi_{d}(\phi)/\phi_{d0}$,
where $\phi_{d0}=\phi_{d}(\phi)|_{\phi=0}$. 
\begin{figure}
\begin{centering}
\includegraphics[width=0.5\textwidth]{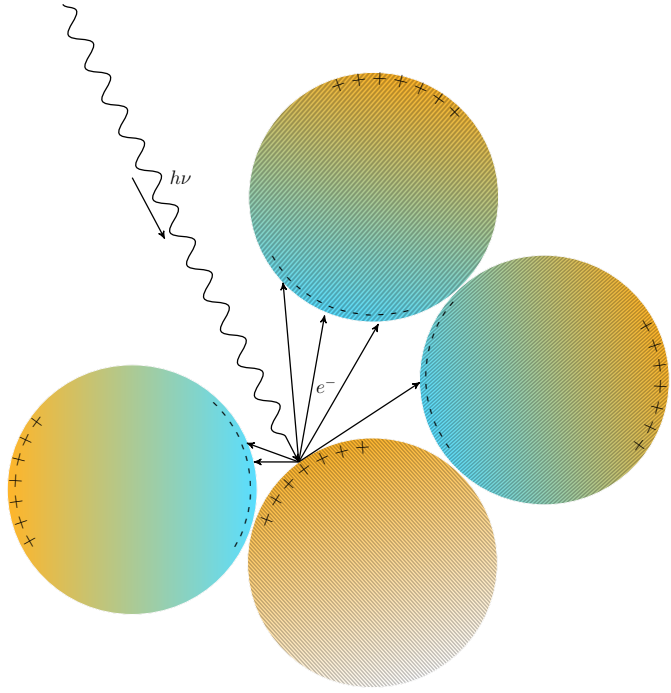}
\par\end{centering}
\caption{\label{fig:A-schematic-representation}A schematic representation
of the Patched Charge Model.}
\end{figure}

\subsection{Inverse sheath and the Patched Charge Model}

As has been already mentioned, we note that there are phenomena related
to dust in astrophysical environments, which are not adequately explained
by classical dust charging models. One of the newest model which has
been developed is the \emph{Patched Charge Model }(PCM) \citep{Wang,Schwan,19},
which can account for existence of dust particles with large negative
charge in an environment which support\textcolor{red}{s} emission
and absorption of photoelectrons.

In the usual case, it is customary to assume that the dust particles
be positively charged due to emission of photoelectrons following
the same procedure which charges the lunar regolith to a positive
potential on the dayside, resulting the existence of an inverse sheath.
However, the PCM can explain why, even in such an environment, the
dust particles on such regolith surfaces\textcolor{red}{{} }may attain
large negative potential. In the PCM, there are formation of \emph{micro
cavities} created by the voids between the dust particles which can
create large negative electric field due to absorption of photoelectrons
(which are emitted by a neighboring dust particle) by the \emph{inner}
surface of the surrounding dust particles (see Fig.\ref{fig:A-schematic-representation}).
The basic principle of the PCM has already been verified through various
laboratory experiments \citep{Wang}, which can be used to investigate
the phenomena of levitation of negatively charged dust particles in
the presence of an inverse sheath. The emission of photoelectrons
by the \emph{inner} particles in the micro cavities and absorption
of these photoelectrons by the dust particles lying on the surface,
may charge the dust particles on the surface to enormously high net
negative potentials despite their \emph{outer }surface being charged
to a positive potential by the incoming UV photons. The laboratory
experiments, simulating the lunar regolith surface shows that this
inter-particle negative electric field is enough to cause dust mobilization
and lofting up to about $0.11\,{\rm m}$ above the surface \citep{Wang}.
We further argue that once the initial mobilization is effected causing
lofting of the dust particles above the surface, the ensuing ion-drag
force will be able to sustain the levitation and push it further up
until the net force becomes zero and a stable levitation height is
obtained. While considering this scenario, we assume that the lofted
dust particles will have the same amount of net negative charge, which
they accumulated at the regolith surface. In what follows, we shall
assume that the dust particles are negatively charged despite being
present in a photoelectron-rich environment.

\subsection{Forces on the dust particles}

In this section, we calculate the forces on a dust particle assume
it to be negatively charged. The primary forces which act on a dust
particle immersed in a plasma are electrostatic $(\bm{F}_{E})$, gravity
$(\bm{F}_{g})$, polarization $(\bm{F}_{{\rm pol}})$, ion drag $(\bm{F}_{{\rm ion}})$,
and neutral drag $(\bm{F}_{{\rm n}})$ force, so that the total force
$\bm{F}$ on a dust particle is given by \citep{Shukla},
\begin{equation}
\bm{F}=\bm{F}_{E}+\bm{F}_{{\rm pol}}+\bm{F}_{g}+\bm{F}_{{\rm ion}}+\bm{F}_{{\rm n}}.
\end{equation}
In what follows, we shall neglect $\bm{F}_{{\rm n},{\rm pol}}$ as
the neutral drag force is proportional to the dust thermal velocity
$u_{d}$ which is $\ll u_{i}$ and so in the ion-acoustic time scale,
it can be safely neglected. Besides the polarization force is important
\emph{only }except in highly dense dusty plasmas and can be neglected.
The normalization factor for force $\bm{F}$ is conveniently expressed
as $(4/3)\pi\lambda_{D}^{3}\rho_{d}g$, which is the total amount
of gravitational force acting on dust particles, those are within
a Debye sphere, $\rho_{d}$ being the matter density of dust particles.
The normalized expressions for $\bm{F}_{E,g}$ are given by,
\begin{eqnarray}
\bm{F}_{E} & = & -3r_{d}R\phi_{d}\left(\frac{d\phi}{dx}\right),\\
\bm{F}_{g} & = & -r_{d}^{3},
\end{eqnarray}
where $R$ is a dimensionless parameter which represents the ratio
of the thermal force on the electrons $(F_{p}=n_{0}T_{e}/\lambda_{D})$
to the gravitational force $(F_{g}=\rho_{d}g)$ on the dust particles,
\begin{equation}
R=\frac{F_{p}}{F_{g}}=\frac{n_{0}T_{e}/\lambda_{D}}{\rho_{d}g}.
\end{equation}
Note that the gravity acts downward \emph{into} the sheath. The ion
drag force is actually due to two factors --- momentum transfer due
to ion-dust collision (collection force $\bm{F}_{{\rm coll}}$) and
Coulomb scattering part of ion-dust collision ($\bm{F}_{{\rm Coul}}$),
which is given by (dimensional),
\begin{eqnarray}
\bm{F}_{{\rm ion}} & = & \bm{F}_{{\rm coll}}+\bm{F}_{{\rm Coul}},\\
\bm{F}_{{\rm coll}} & = & \pi b_{{\rm max}}^{2}m_{i}n_{i}\bar{u}\bm{u}_{i},\qquad\bar{u}=\left(u_{i}^{2}+u_{s}^{2}\right)^{1/2},\\
\bm{F}_{{\rm Coul}} & = & 4\pi b_{\perp}^{2}m_{i}n_{i}\bar{u}\bm{u}_{i}\,\ln\left(\frac{\lambda_{D}^{2}+b_{\perp}^{2}}{b_{{\rm max}}^{2}+b_{\perp}^{2}}\right)^{1/2},
\end{eqnarray}
where
\begin{equation}
b_{{\rm max}}=r_{d}\left(1-\frac{2e\phi_{d}}{m_{i}\bar{u}^{2}}\right)^{1/2},\qquad b_{\perp}=\frac{eQ_{d}}{4\pi\epsilon_{0}m_{i}\bar{u}^{2}}.
\end{equation}
Note that $b_{{\rm max}}$ and $b_{\perp}$ have dimensions of length.
The normalized expression for $\bm{F}_{{\rm ion}}$ is given by,
\begin{equation}
\bm{F}_{{\rm ion}}=\hat{\bm{\iota}}\frac{3}{4}R\bar{u}M\left[b_{{\rm max}}^{2}+2b_{\perp}^{2}\ln\left(\frac{1+b_{\perp}^{2}}{b_{{\rm max}}^{2}+b_{\perp}^{2}}\right)\right],
\end{equation}
where $\hat{\bm{\iota}}$ is the unit vector in the direction of $\bm{u}_{i}$.

\begin{figure}
\begin{centering}
\includegraphics[width=0.5\textwidth]{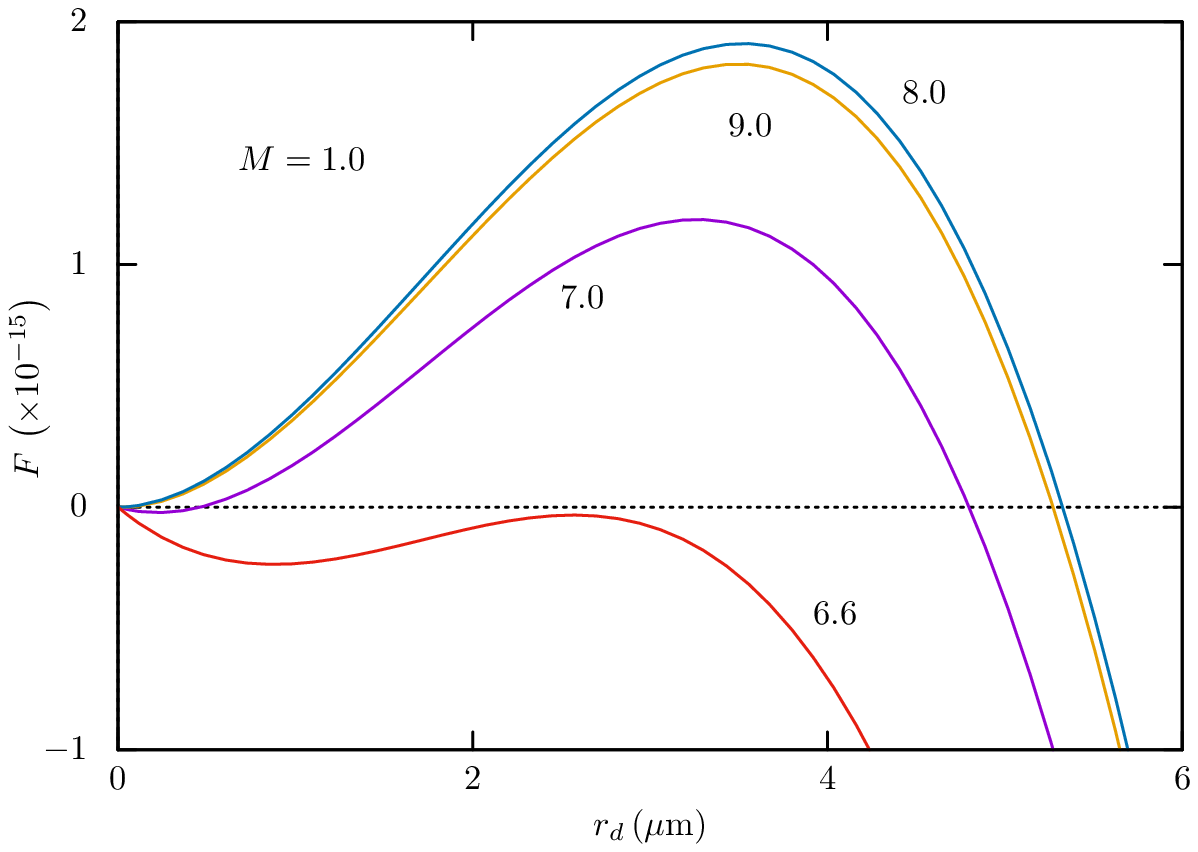}\hfill{}\includegraphics[width=0.5\textwidth]{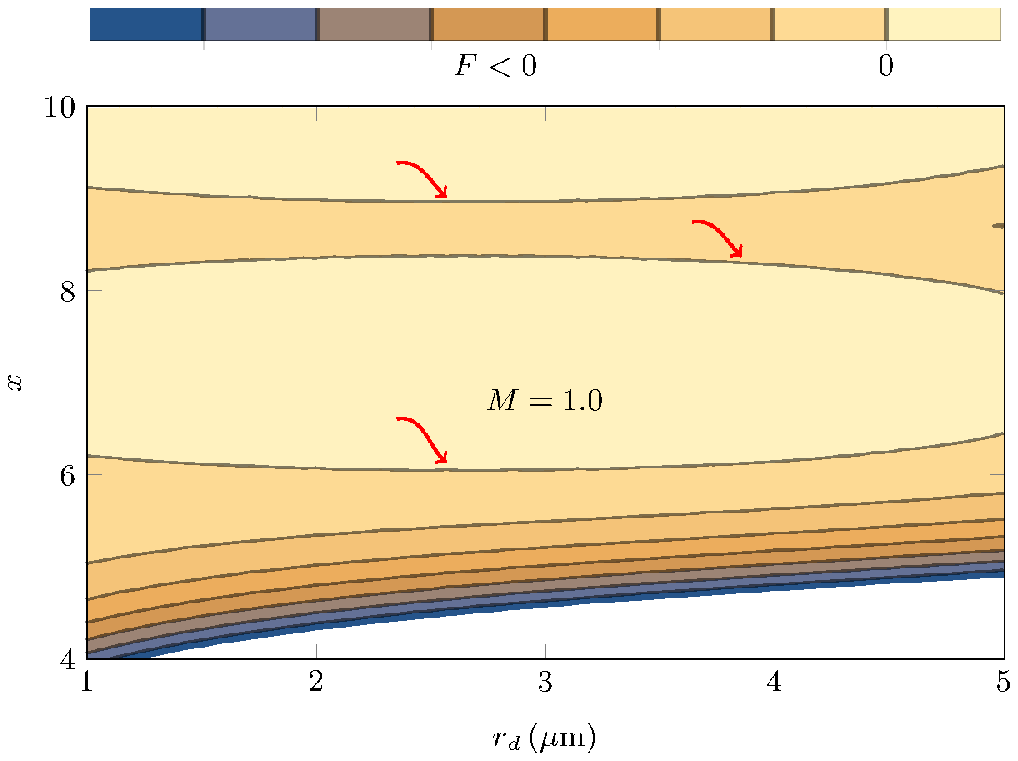}\\
\includegraphics[width=0.5\textwidth]{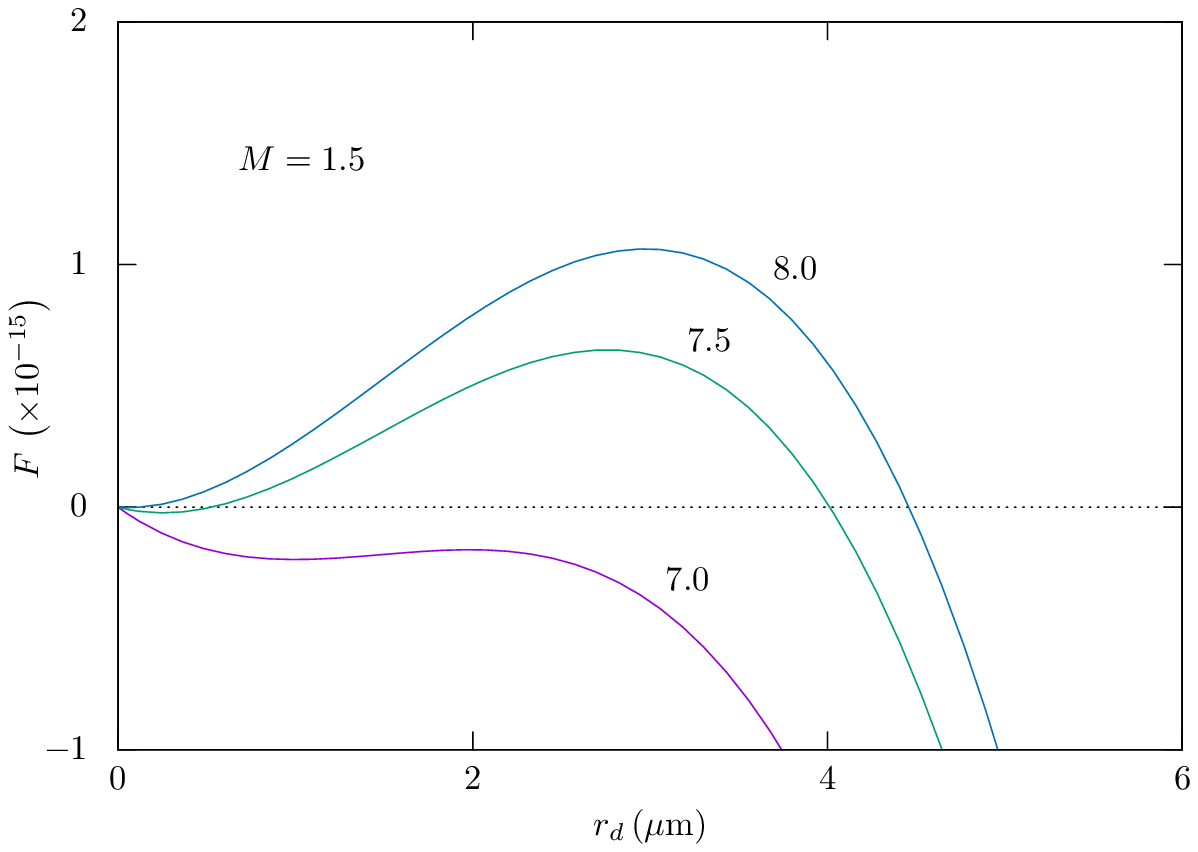}\hfill{}\includegraphics[width=0.5\textwidth]{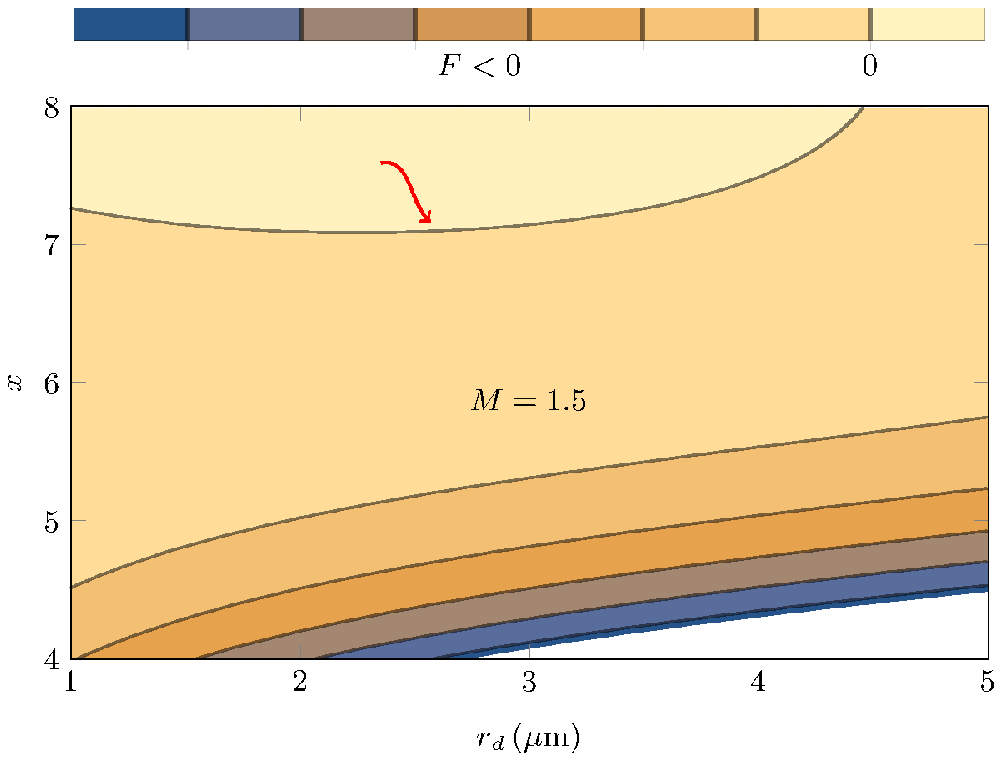}
\par\end{centering}
\caption{\label{fig:The-levitation-of}The levitation of dust particles above
an inverse sheath for two Mach numbers $1.0$ and $1.5$. The left
panels show the total force experienced by the dust particles. The
inscribed numbers on the curves denote values of $x$. The parameters
used here are for lunar surface. $n_{e0}=10^{12}\,{\rm m}^{-3},T_{e}=1\,{\rm keV},J_{h\nu}=4.5\,\mu{\rm A/m},\sigma=1,\delta_{i}=1.1,\sigma_{{\rm ph}}=1,\delta_{w}=1.5$.
The right panels show the contours of total force $\bm{F}$ for the
shown parameter space. The red arrows indicate the lines where $\bm{F}=0$.}
\end{figure}

As the inverse sheath is charged to a positive potential, the dust
particles feel a downward (toward the sheath) pull due to gravity
and the electric field, whereas the ion drag force acts away from
the sheath. In Fig.\ref{fig:The-levitation-of}, we show the results
of our numerical calculations for lunar plasma parameters --- $T_{e}=1\,{\rm keV}$
\citep{Horanyi}, $n_{e0}=10^{12}\,{\rm m^{-3}}$ \citep{Popel},
$J_{h\nu}=4.5\,\mu{\rm A/m^{2}}$ \citep{Feuerbacher}, and $\rho_{d}=1000\,{\rm kg/m^{3}}$
\citep{5} for two Mach numbers $M=1.0$ and $1.5$. We note here
that there are also reports about much higher photoelectron density
of $\sim2\times10^{14}\,{\rm m^{-3}}$ above the dayside lunar surface
\citep{Burinskaya}. The panels on the left show the net force experienced
by a dust particle which initially tends to be slight negative (downward,
toward the sheath) for smaller dust particles. For larger dust particles,
the net force is positive (upward, away from the sheath). For even
larger dust particles, the net force again becomes negative pulling
the massive dust particles down to the surface. This behavior can
be understood on the basis of the fact that the accumulation of negative
charges on a dust particles is proportional to its size, which increases
the ion drag force and is responsible for levitation of the dust particles.
However, as dust size grows, the force due to gravity becomes dominant
as well as the electrostatic attractive force, which eventually win
for the larger dust particles. The panels on the right show the contours
of the net force $\bm{F}$ experienced by the dust particles in the
$(r_{d},x)$ parameter space, where the red arrows indicate the lines
for $\bm{F}=0$. What is interesting and a unique behaviour that we
see is that formation of \emph{dust-bands} (position where $\bm{F}=0$)
or \emph{striations }at different heights as one moves upward from
lunar surface, which is at $x=0$. The bands or striations are caused
by the profile of the ion-drag force which has a minima as one comes
away from the surface (see Fig.\ref{fig:The-electric-field}). In
terms of physical distance, the bands (as seen in Fig.\ref{fig:The-levitation-of})
are formed at about $\sim1.4,1.9,$ and $2.1\,{\rm m}$ from the lunar
surface, with the Debye shielding length $\sim0.23\,{\rm m}$ for
the assumed lunar plasma parameters, which compares very well with
the reported levitation height of $\sim2\,{\rm m}$ for dust particles
above the lunar surface \citep{Hatzell} with an average size of $\sim2-4\,{\rm \mu m}$
for the dust particles \citep{Horanyi}. The most important and significant
finding of our calculations is that the levitating height of the dust
particles remains same for a considerable size distributions of the
dust particles, which is quite opposite of what is found in case of
classical dust-levitation scenario with only electrostatic and gravitational
force \citep{gcdas}. As can be seen from Fig.\ref{fig:The-levitation-of}
(upper panel), the levitation height remains reasonably constant for
dust sizes from about $1.5\,{\rm \mu m}$ to $4.5\,{\rm \mu m}$ for
a Mach number of $1.0$. This property of the dust-levitation causes
formation of bands or striations and may have explanation to lunar
phenomena such as the horizon glow. 

We would like to note that the solar wind plasma has a thermal velocity
of $\sim300\,{\rm km/s}$ with a bulk plasma velocity of about $\sim400\,{\rm km/s}$
\citep{McComas}, which is toward the dayside lunar surface. This
velocity is incorporated into the calculation through the Mach number,
which is the velocity of the solar wind ions far away from the sheath.
So, a Mach number of $1.2$ results a bulk ion velocity of $\sim380\,{\rm km/s}$
and a velocity of $\sim450\,{\rm km/s}$ is equivalent to Mach number
of $\sim1.4$. We have shown the formation of levitated dust bands
even for a higher Mach number of $M=1.5$.

\subsubsection{An estimate of the ion-drag force}

\begin{figure}
\begin{centering}
\includegraphics[width=0.5\textwidth]{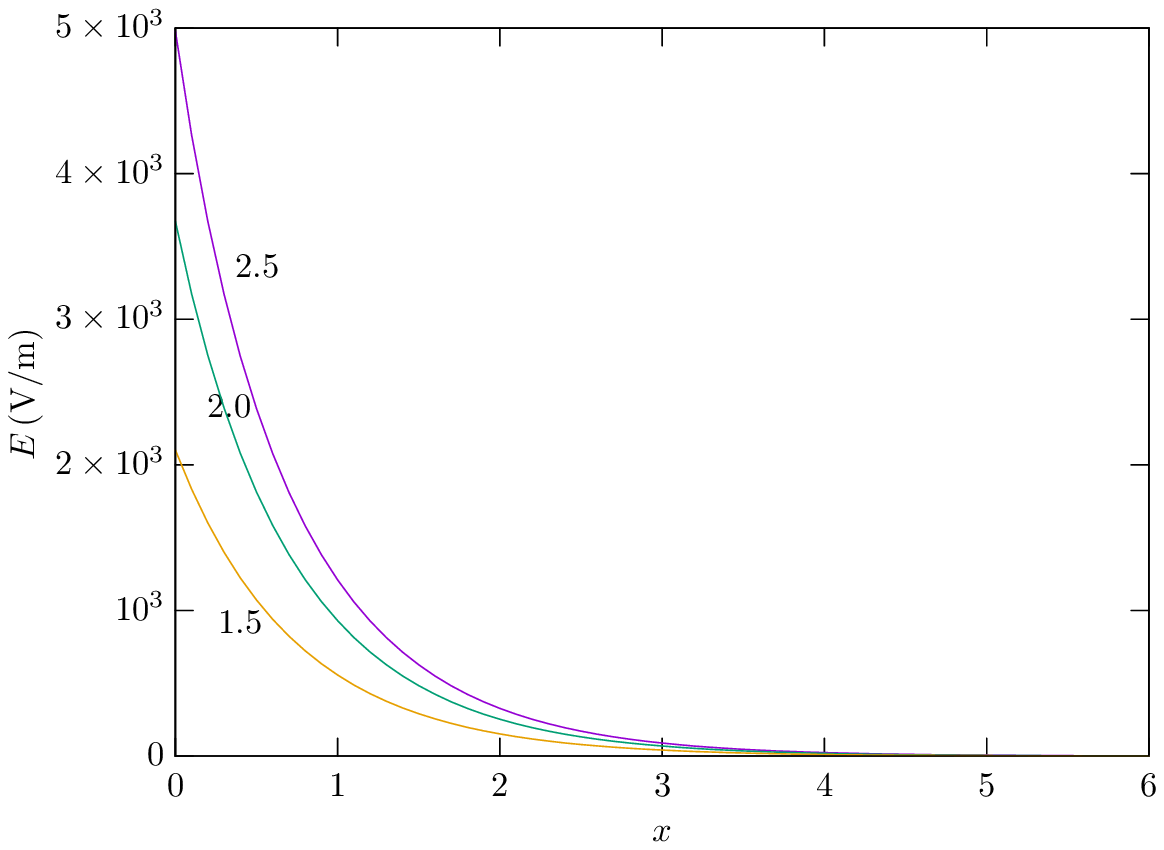}\hfill{}\includegraphics[width=0.5\textwidth]{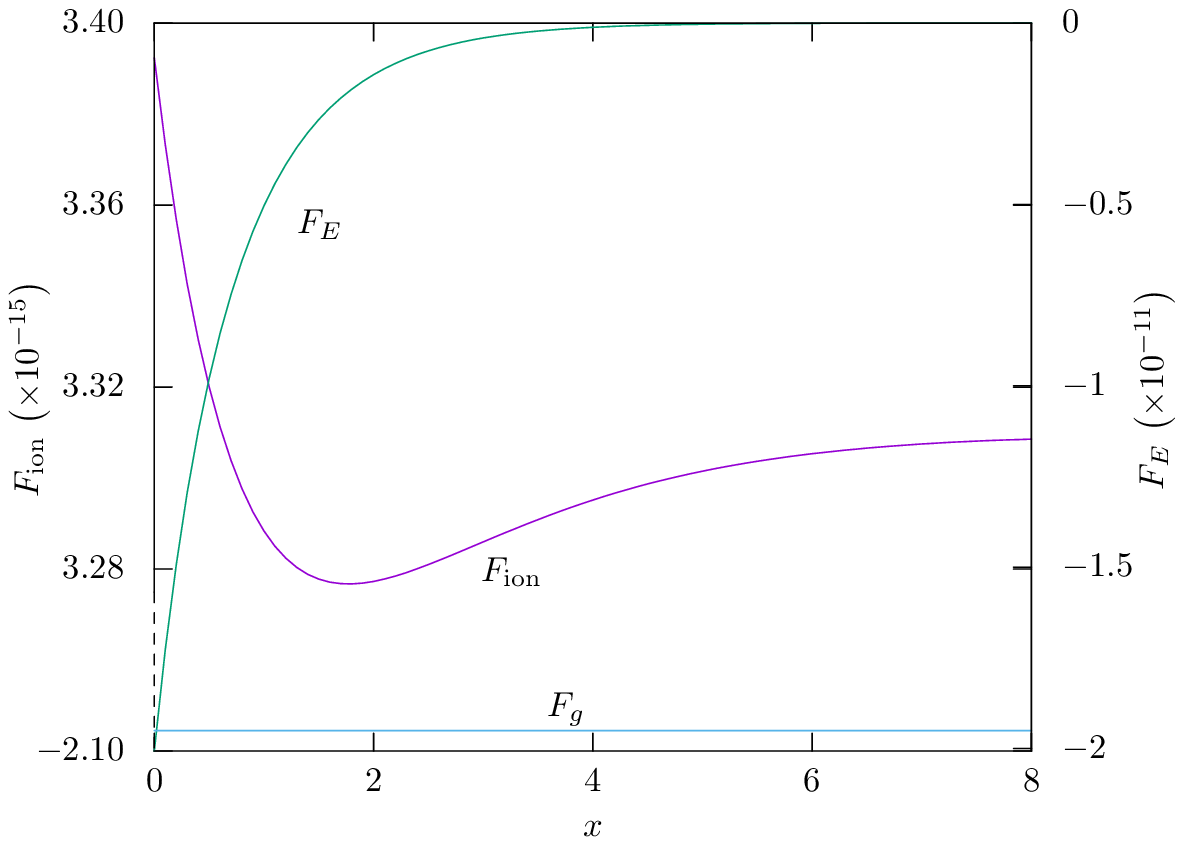}
\par\end{centering}
\caption{\label{fig:The-electric-field}The electric field (left) is shown
for the parameters $n_{e0}=10^{7}\,{\rm m}^{-3},T_{e}=1\,{\rm keV},J_{h\nu}=4.5\,\mu{\rm A/m^{2}},M=1.2,\sigma=1,\delta_{i}=1.1,\sigma_{{\rm ph}}=1$.
The numbers in the figure indicates $\delta_{{\rm ph}}$. The panel
on the right shows the magnitudes of the three forces (normalized)
for the same parameters and the dust particle size $3\times10^{-6}\,{\rm m}$.}
\end{figure}

We now try to calculate the forces at the wall, $x=0$, where $x$
is the distance from the wall measured in the unit of $\lambda_{D}$.
For dust matter density of $\sim1000\,{\rm kg/m^{3}}$, average dust
radius of $\sim10^{-6}\,{\rm m}$, $T_{i}\sim T_{e}\sim1\,{\rm keV}$,
$n_{0}\sim10^{12}\,{\rm m^{3}}$, we find that
\begin{equation}
b_{{\rm max}}\simeq7\times10^{-5}\quad{\rm and}\quad\bar{u}\sim1.7.
\end{equation}
As such, the expression for $F_{{\rm ion}}$ becomes
\begin{equation}
F_{{\rm ion}}\sim\frac{3}{4}R\bar{u}Mb_{{\rm max}}^{2}.
\end{equation}
For Mach no $\sim1$, we see that $F_{{\rm ion}}\sim4.8\times10^{-16}$
(normalized). A simple calculation shows that the force due to gravity
$F_{g}\sim7.7\times10^{-17}$, and that due to the electric field
is $F_{E}\sim6.8\times10^{-12}$. However, as one moves away from
the surface, the force due to electric field on dust particles decreases
as the dust potential becomes less negative (see Fig.\ref{fig:The-electric-field}).
Note that the dust potential is self-consistently calculated using
the current balance condition (Eq.\ref{eq:dcharge} in the manuscript).
In Fig.\ref{fig:The-electric-field}, we have shown variation of the
relative magnitudes of the forces with the distance from the wall.
Clearly, as we move away from the surface, the the ion-drag force
dominates and becomes capable of balancing the other two forces leading
to dust levitation. 

In this context, we would like to draw the attention to a very recent
experimental work on ion-drag force on dust particles in laboratory
plasmas, where the ion-drag force is found to be almost of the same
order as the force due to the electric field and the force due to
gravity (i.e. earth gravity) for similar plasma parameters \citep{Bailung}.
We also note that the plasma parameters used in this work are also
very well realizable in laboratory conditions.

\section{Summary and Conclusion}

In summary, a self consistent analytical model describing the inverse
sheath structure over the lunar surface has been developed. We have
presented the basic formulation for one dimensional collisionless
inverse plasma sheath consisting of electrons and ions with considerable
presence of dust particles. The electrons have two populations ---
the plasma electrons and the photoelectrons emitted by the wall. In
our case, the plasma electron is described by a single population
and photoelectron density is described by two populations --- one
which is sheath limited population and another which contribute to
the bulk plasma electrons. The electron densities have been derived
by assuming both the plasma electrons and photoelectrons populations
to be Maxwellian. In deriving the sheath structure and dust levitation
phenomena, a steady state potential over the lunar regolith has been
determined. The dust charge has been consistently determined by balancing
the currents to the surface of the dust particles. Our analysis has
been found to be relevant in both laboratory and space plasma environments.

When the lunar surface is exposed to the solar UV radiation and solar
wind plasma, the electrostatic charging of surface as well as the
dust particles can lead to dust levitation under favourable circumstances.
Previous charging models were unable to account for particle charges
large enough to attain the kinds of motion observed either in space
or in lab conditions \citep{Wang1,Flanagan}. Laboratory based experiments
have supported the \textquotedblleft Patched Charge Model\textquotedblright ,
which can explain why the dust particles lying on a regolith surface
may attain large negative charge, even in a photoelectron-rich environment
under favourable conditions. According to this model, micro cavities
form between dust particles and the shielded surface patches amass
large amounts of negative charge from photoelectrons and secondary
electrons emitted by the surface layer patches with direct UV or electron
beam exposure. The collected negative charge may become large enough
to cause dust mobilization and dust lofting. We have argued that the
ion drag force can sustain the levitation of negatively charged dust
particles in an inverse sheath, once their initial lofting is effected
due to large negative charge the dust particles may acquire following
formation of micro cavities on the surface. Our analysis show that
levitation of the negatively charged dust particles is possible in
the inverse photoelectron sheath under favourable parameter space
owing to the ion drag force on the dust particles. Though in most
of the cases, the ion-drag force remains negligible, it can play an
important role in suitable parameter regime as shown in this work
and also in laboratory environments. 

We have analytically determined the heights of the levitating dust
grains from the surface of the moon with lunar plasma parameters.
In our study, we have examined the levitation distance from the surface
for different sizes of dust particles. In certain cases, there are
more than one levitation heights leading to the formation of bands
of dust particles above the surface. One important finding of our
work is that the levitation heights of these negatively charged dust
particles remain almost same for dust particles of different sizes,
ranging from about $1.5\,{\rm \mu m}$ to $4.5\,{\rm \mu m}$, which
support the formation of bands of dust particles over the lunar surface.

\section*{Acknowledgement}

It is a pleasure to thank the anonymous referee for a critical review
of the manuscript. One of the authors, RD thanks UGC, India for the
BSR Fellowship during which the work has been carried out.

\bibliographystyle{phaip}
\bibliography{ref}

\begin{thebibliography}{10}

\bibitem{1}
K.~U. Riemann,
\newblock Journal of Physics D {\bf 24}, 493 (1991).

\bibitem{2}
X.~Wang, J.~Pilewskie, H.-W. Hsu, and M.~Hor\'anyi,
\newblock Geophysical Research Letters {\bf 43}, 525 (2016).

\bibitem{3}
R.~H. Manka,
\newblock in {\em Photon and Particle Interactions with Surfaces in Space},
  edited by R.~J.~L. Grard, pages 347--361, Dordrecht, Reidel, 1973.

\bibitem{4}
G.~D. Hobbs and J.~A. Wesson,
\newblock Journal of Plasma Physics {\bf 9}, 85 (1967).

\bibitem{5}
T.~O. Nitter, O.~Havnes, and F.~Melands\o,
\newblock Journal of Geophysical Research {\bf 103}, 6605 (1998).

\bibitem{6}
A.~Poppe and M.~Hor\'anyi,
\newblock Journal of Geophysical Research {\bf 115}, A08106 (2010).

\bibitem{7}
A.~R. Poppe, M.~Piquette, L.~A., and M.~Hor\'anyi,
\newblock Icarus {\bf 221}, 135 (2012).

\bibitem{Bailung}
Y.~Bailung et~al.,
\newblock Physics of Plasmas {\bf 25}, 053705 (2018).

\bibitem{8}
A.~Dove et~al.,
\newblock Physics of Plasmas {\bf 19}, 043502 (2012).

\bibitem{9}
J.~P. Sheehan et~al.,
\newblock Physical Review Letters {\bf 111}, 75002 (2013).

\bibitem{10}
S.~Langendorf and M.~Walker,
\newblock Physics of Plasmas {\bf 22}, 33515 (2015).

\bibitem{11}
S.~F. Singer and E.~H. Walker,
\newblock Icarus {\bf 1}, 112 (1962).

\bibitem{12}
J.~J. Rennilson and D.~R. Criswell,
\newblock The Moon {\bf 10}, 121 (1974).

\bibitem{13}
M.~A. Pelizzari and D.~R. Criswell,
\newblock in {\em 9th Lunar Planetary and Science Conference}, pages
  3225--3237, 1978.

\bibitem{14}
H.~A. Zook and J.~E. McCoy,
\newblock Geophysical Research Letters {\bf 18}, 2117 (1991).

\bibitem{lunar_prospector}
A.~B. Binder,
\newblock Science {\bf 281}, 1475 (1998).

\bibitem{apollo1}
O.~E. Berg, H.~Wolf, and J.~Rhee,
\newblock in {\em Interplanetary Dust and Zodiacal Light}, edited by H.~Elssser
  and H.~Fetching, page 233, Springer-Verlag, New York, 1976.

\bibitem{apollo2}
P.~D. Feldman and D.~Morrison,
\newblock Gephysical Research Letter {\bf 18}, 2105 (1991).

\bibitem{apollo3}
O.~E. Berg, F.~F. Richardson, and H.~Button,
\newblock in {\em Apollo 17 Preliminary Science Report}, NASA Space. Publ.
  SP-330, 16-1-16-9, 1973.

\bibitem{15}
J.~S. Halekas, Y.~Saito, D.~G. T., and F.~W. M.,
\newblock Planetary and Space Science {\bf 59}, 1681 (2011).

\bibitem{gcdas}
G.~C. Das, R.~Deka, and M.~P. Bora,
\newblock Physics of Plasmas {\bf 23}, 042308 (2016).

\bibitem{16}
S.~K. Mishra and S.~Misra,
\newblock Physics of Plasmas {\bf 25}, 023702 (2018).

\bibitem{Wang}
X.~Wang, J.~Schwan, H.-W. Hsu, E.~Gr{\"u}m, and M.~Hor{\'a}nyi,
\newblock Geophysical Research Letters {\bf 43}, 6103 (2016).

\bibitem{19}
S.~Joseph,
\newblock Undergraduate Honors Theses, 1442, University of Colorado at Boulder,
  2017.

\bibitem{17}
M.~Hor\'anyi,
\newblock Annual Review of Astronomy and Astrophysics {\bf 34}, 383 (1996).

\bibitem{18}
H.~A. Zook, A.~E. Potter, and B.~L. Cooper,
\newblock in {\em Lunar and Planetary Science Conference}, volume~26, pages
  1577--1578, 1995.

\bibitem{Sagdeev}
R.~Z. Sagdeev,
\newblock in {\em Reviews of Plasma Physics}, edited by M.~A. Leontovich,
  volume~4, page~23, Consultants Bureau, New York, 1966.

\bibitem{Shukla}
P.~K. Shukla and A.~A. Mamun,
\newblock Institute of Physics, Bristol and Philadelphia, 2002.

\bibitem{Feuerbacher}
B.~M. Feuerbacher, B.~F. Anderegg, L.~D. Laude, R.~F. Willis, and R.~J.~L.
  Grard,
\newblock in {\em Lunar Science Conference}, volume IIIrd, pages 2655--2663,
  1972.

\bibitem{Horanyi}
M.~Hor\'anyi, B.~Walch, S.~Robertson, and D.~Alwxander,
\newblock Journal of Gephysical Research {\bf 103}, 8575 (1998).

\bibitem{Popel}
S.~I. Popel et~al.,
\newblock JETP Letters {\bf 99}, 115 (2014).

\bibitem{Burinskaya}
T.~M. Burinskaya,
\newblock Planetary and Space Science {\bf 115}, 64 (2015).

\bibitem{Hatzell}
C.~M. Hatzell and D.~J. Scheeres,
\newblock JGRE {\bf 118}, 116 (2013).

\bibitem{McComas}
D.~J. McComas et~al.,
\newblock Geophysical Research Letters {\bf 30}, 1517 (2003).

\bibitem{Wang1}
X.~Wang, M.~Hor{\'a}nyi, and S.~Robertson,
\newblock Planetary and Space Science {\bf 59}, 1791 (2011).

\bibitem{Flanagan}
T.~M. Flanagan and J.~Groree,
\newblock Physics of Plasmas {\bf 13}, 123504 (2006).

\end{thebibliography}

\end{document}